\documentclass[aps,prd,10pt,onecolumn,nofootinbib,showpacs,showkeys,notitlepage]{revtex4-1}

\usepackage[dvipdfmx]{graphicx} 
\usepackage{bmpsize}

\usepackage{amsmath}
\usepackage{caption}
\usepackage{subcaption}
\usepackage{slashed}
\usepackage{color}
\usepackage{soul}
\usepackage{amsfonts}

\newcommand{\beq}{\begin{equation}}
\newcommand{\eeq}{\end{equation}}
\newcommand{\bea}{\begin{eqnarray}}
\newcommand{\eea}{\end{eqnarray}}

\newcommand{\nn}{\nonumber\\}

\renewcommand{\d}{\partial}

\newcommand{\x}{{\bf x}}

\newcommand{\rh}{\varrho}

\newcommand{\exv}[1]{\left\langle{#1}\right\rangle}

\newcommand{\ep}{\varepsilon}

\newcommand{\sgn}{\mathop{\textrm{sgn}}}

\newcommand{\pint}[2]{{\int\!\frac{d^{#1}#2}{(2\pi)^#1}\,}}

\newcommand \be {\begin{equation}}
\newcommand \ee {\end{equation}}
\newcommand \bed {\begin{displaymath}}
\newcommand \eed {\end{displaymath}}

\newcommand{\bit}{\begin{itemize}}
\newcommand{\eit}{\end{itemize}}

\newcommand{\la}{\leftarrow}
\newcommand{\ra}{\rightarrow}

\def\s0#1#2{\mbox{\small{$ \frac{#1}{#2} $}}}
\def\0#1#2{\frac{#1}{#2}}

\def\la{\langle}
\def\ra{\rangle}

\begin{document}

\title{Infrared multiphoton resummation in quantum electrodynamics} 
\author{P. Mati}

\affiliation{ELI-ALPS, ELI-Hu NKft, Dugonics t\'er 13, Szeged 6720, Hungary}
\email{Peter.Mati@eli-alps.hu, matipeti@gmail.com} 

\begin{abstract}

Infrared singularities in massless gauge theories are known since the foundation of quantum field theories. The root of this 
problem can be tracked back to the very definition of these long-range interacting theories such as QED. It can be shown that 
singularities are caused by the massless degrees of freedom (i.e. the photons in the case of QED). In the Bloch-Nordsieck 
model the absence of the infrared catastrophe can be shown exactly by the complete summation of the radiative corrections. In 
this paper we will give the idea of the derivation of the Bloch-Nordsieck propagator, that describes the infrared structure of the 
electron propagation, at zero and finite temperatures. Some ideas of the possible applications are also mentioned.

\end{abstract}

\maketitle

\section{Introduction}
The infrared (IR) limit of quantum electrodynamics (QED) is known to be plagued by singularities caused by the 
photons. This phenomenon is known as the infrared catastrophe, and it can be found in any quantum field 
theory (QFT) which involves massless fields. The development of QFTs started around 1930 with QED, therefore, 
in most of the cases the subjects of the computations were electromagnetic quantities. The methods used for the 
calculations were mostly the direct extension of the PT from quantum mechanics. 
Physicists back then, who were doing computations in QED, immediately faced IR divergences when 
calculating first order perturbative corrections to the Bremsstrahlung process, due to the low frequency photon 
contributions. The core of the problem lays in the fundamental definition of QED, namely, that we assume the 
existence of a free theory, i.e. the existence of asymptotic states. However, such states are difficult to define in a 
theory where we have long-range interactions. As a consequence, one cannot truly define the asymptotic states 
described by the Fock representation of free theory Hilbert space, on which the PT is performed. Thus, we need to 
search for a non-perturbative solution to prevent these difficulties. An alternative approach to this problem was 
provided by Bloch and 
Nordsieck in 1937 in their remarkable work on treating the infrared problem \cite{BN}. The divergencies are 
caused by the fact that in a scattering process an infinite amount of long wavelength photons are emitted, and 
these low energy excitations of the photon field are always present around the electron in the form of a "photon 
cloud". This shows us essentially that the observed particle is in fact very different from the one we call the bare 
particle: they can be considered as dressed "quasi particle" objects whose interactions cannot be described 
through PT entirely.
In this paper, we will show the emergence of the infrared catastrophe and then we will introduce the Bloch-Nordsieck (BN) 
model, which was designed in order to imitate the low energy regime of QED. We will discuss the 
breakdown of the PT due to the IR catastrophe, however, it is possible to obtain the exact full solution by using the 
Ward-Takahashi identities embedded into the Dyson-Schwinger (DS) equation. 

\section{The infrared catastrophe}
The easiest way to demonstrate the IR catastrophe is the following. Suppose that $E(\omega)$ is a finite 
amount of electromagnetic energy emitted by an accelerated charge in the frequency band $[\omega,\omega+d
\omega]$. Each photon carries an energy of $\hbar\omega$, hence the average number of emitted photons in this 
band is $\bar n = E(\omega)/\hbar \omega $. If we take the limit $\omega\to0$ the average number of 
photons will diverge provided that $\lim_{\omega\to0} E(\omega)\neq0$ (which is fulfilled). Thus we can see that an infinite number of soft photons are present at any scattering process (c.f. \cite{Pes,itzy,Fried}). In the following we will apply the convention used by the particle physics community, i.e. $\hbar=c=1$ in 
the further computations.\\
We should get the same result using quantum computations, however, relying on PT gives different result: 
already in the first 
order of the PT the probability of emitting one soft photon in a scattering process will diverge logarithmically \cite{Pes,itzy}. 
This is in complete contradiction that we have found with the semi-classical line of thought a few lines above. In fact, it turns out it is not enough to take 
into account the tree level diagrams 
but we will also need to include the virtual corrections to cancel out these infrared divergencies. This can be done in 
all orders of PT and by summing up these corrections (combining real and virtual corrections) to infinite order we can obtain a 
well defined probability measure. More precisely, it can be shown (c.f. \cite{Fried}) that the probability of emitting $n$ soft photons 
in the process has the form
\beq\label{nfoton}
P_{\Delta E,\omega_{min}}(n)=\left|\la p' \left| S \right| p \ra\right|^2\frac{1}{n!}\left[\frac{2\alpha}{3\pi}\frac{-q^2}
{m^2}\ln{\frac{\Delta E}{\omega_{min}}} \right]^n e^{-\frac{2\alpha}{3\pi}\frac{q^2}{m^2}\ln{\frac{\Delta E}
{\omega_{min}}}}.
\eeq
Here the first factor is the absolute square of the scattering amplitude without radiation emitted, $\alpha$ is the fine 
structure constant ($\alpha=e^2/4\pi$, with $e$ being the electric charge); $m^2$ and $q^2$ is the electron mass and 
the transferred four momentum $q=p-p'$ (and $-q^2>0$), respectively. There are two energy scales that have been introduced: 
$\omega_{min}$ and $\Delta E$. The former is an artificial IR regulator ("mass" for the photon field) and the second is the 
resolution of the detector that performs the measuring in the process: photons having energy lower than $\Delta E$ 
are not being detected at all. Hence, we only consider the interval where the photons energy are $\omega \in 
[\omega_{min}, \Delta E]$. However, \eqref{nfoton} will give 0 for any finite $n$ while taking the artificial mass of the 
photon to zero as we should:
\beq
\lim_{\omega_{min}\to0}P_{\Delta E,\omega_{min}}(n)=0.
\eeq
This means that the probability of emitting any finite number of soft photons during the scattering process is zero. On the 
contrary: if we perform a summation over all possible photon numbers that can be emitted we will get a finite result
\footnote{This result can be found in \cite{Fried} where functional techniques are used, however, in \cite{Pes} the following formula is given for the differential cross section:
\beq
\frac{d\sigma}{d \Omega}=\left(\frac{d\sigma}{d \Omega}\right)_0 e^{-\frac{\alpha}{\pi}\ln{\frac{-q^2}{m^2}}\ln{\frac{-q^2}{\Delta E^2}}}.
\eeq
Here the first factor corresponds to the hard scattering and in the exponent we can find the famous {\emph {Sudakov double logarithm}}. The difference between the results in \cite{Pes} and \cite{Fried} originates from the different approximations that are used.}
\beq
P(\Delta E)=\sum\limits_{n=0}^{\infty} P_{\Delta E}(n)=\left|\la p' \left| S \right| p \ra\right|^2 e^{-\frac{2\alpha}{\pi}
\frac{-q^2}{m^2}\ln{\frac{m}{\Delta E}}}.
\eeq
As we can see, indeed, we obtained a finite probability for the process of infinite emitted photons from the quantum 
computation. However, we still need to keep the sensitivity of the detector finite in order to get this result. Although 
theoretically it is possible to take the limit $\Delta E\to 0$, however, in reality it will never happen since there is no 
such as a detector 
with perfect resolution.\\

\section{The soft photon contribution to the electron structure: the Bloch-Nordsieck model}

The BN model was made to give an insight in the analytic structure of the cancellation of the 
infinities in the IR regime. Investigating this model will lead us to the exact result of the propagator of the fermion 
which is surrounded by a cloud of photons. Although the 
result is known long ago \cite{BN}, and it can be considered as a textbook material \cite{shirkov}, we will sketch the 
derivation used in \cite{JakoMati1} which can be extended to finite temperatures the most easily \cite{JakoMati2}. The main idea of the 
BN model, that simplifies the theory tremendously, is to replace the gamma matrices $\gamma_{\mu}$ by a four-vector 
$u_{\mu}$ that can be considered as the four-velocity of the fermion, and hence the fermion field is represented by a 
scalar field. This simplification is well justified in the IR regime: the soft photons which take part in the interaction will 
not have enough energy for pair production, moreover, not even enough to flip the spin of the electron. It implies that 
the photon propagator will not have any corrections, i.e. the exact photon propagator is the free one in this model. 
Thus the Lagrangian reads 
\beq
  {\cal L} = -\frac14 F_{\mu\nu} F^{\mu\nu} + \bar \Psi(iu_\mu D^\mu
  - m)\Psi,\qquad iD_\mu = i\d_\mu -eA_\mu,\quad F_{\mu\nu}=\d_\mu
  A_\nu - \d_\nu A_\mu,
\eeq
where $\Psi$ and $A_{\mu}$ is the fermion and photon field, respectively. 
The fermionic part of the Lagrangian is Lorentz-covariant, therefore
we can relate the results with different $u^\mu$ choice by Lorentz
transformation. This makes possible to work with $u=(u_0,0,0,0)$ without
loss of generality. In fact, we can perform a Lorentz-transformation
where $\Lambda u=(u_0,0,0,0)$. Since $u^\mu$ is a four-velocity then
$u_0=1$; if it is of the form $u=(1,\mathbf{v})$, then it is
$u_0=\sqrt{1-\mathbf{v}^2}$. After rescaling the field as $\Psi\to
\Psi/\sqrt{u_0}$ and the mass as $m\to u_0m$, the Lagrangian reads
\begin{equation}
  {\cal L} = -\frac14 F_{\mu\nu} F^{\mu\nu} + \bar \Psi(iD_0-m)\Psi.
\end{equation}
Using this reference frame simplifies the calculations \cite{JakoMati1}. If necessary, the complete $u$ dependence can 
be recovered easily, however, at finite temperatures we need to use numerics if we want to switch to another reference frame 
due to the lack of the Lorentz symmetry.\\

\subsection{The Bloch-Nordsieck model at $T=0$}

\begin{figure}[htbp]
  \centering
  \includegraphics[width=0.35\textwidth]{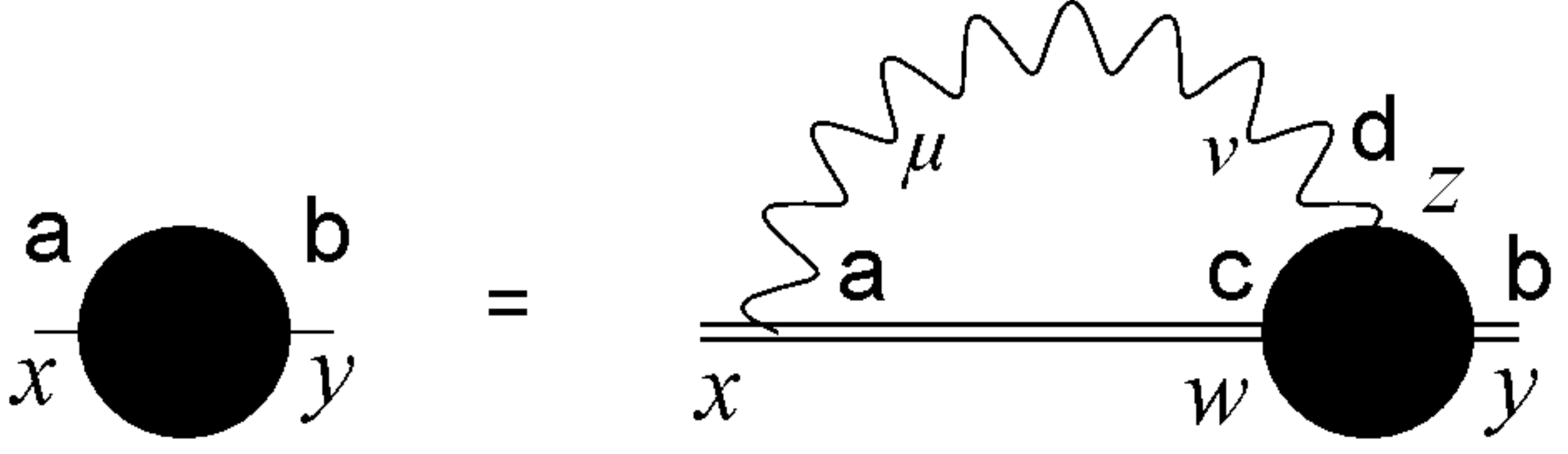}
  \caption{The diagrammatic representation of the Dyson-Schwinger equation. The double line is for the dressed fermion, the wavy line is for the photon 
  propagator. The black blob denotes the full vertex function.  The bold letters are for space-time points 
  and the Greek letters denote the Lorentz indices. At finite temperatures we also need to consider the regular letters 
  which are the Keldysh indices.}
  \label{fig:DS}
\end{figure}

In order to derive the dressed fermionic propagator, we are going to use a system of self-consistent equations which 
involves the DS equation, Dyson's series and the Ward-Takahashi identities. We will begin with 
the DS equation; it reads in real and momentum space, respectively: 
\bea\label{DSeq}
  \Sigma(x-y)&=&-ie^2\int d^4 w \, \int d^4 z \, {\cal G}(x-w) u^\mu G_{\mu\nu}(x-z) \Gamma^\nu(z;w,y),\nn
  \Sigma(p)&=&-ie^2\pint4k {\cal G}(p-k) u^\mu G_{\mu\nu}(k) \Gamma^\nu(k;p-k,p),
\eea
where $G_{\mu\nu}$ is the photon propagator, $\cal G$ is the fermion propagator and $\Gamma^\mu$ is the vertex function. The diagrammatic representation of equation \eqref{DSeq} can be seen in Fig.~\ref{fig:DS}. The DS equation 
describes the self-energy of the fermion. To obtain the full expression, we need to treat $\cal G$ as the exact fermion 
propagator and keep the photon propagator undressed (i.e. on tree-level), which coincides with the exact one in the 
framework of the BN model. The vertex correction is composed from both propagators but 
it can be simplified using the Ward-Takahashi relations, as we will see.\\

\begin{figure}[htbp]
  \centering
  \includegraphics[width=0.7\textwidth]{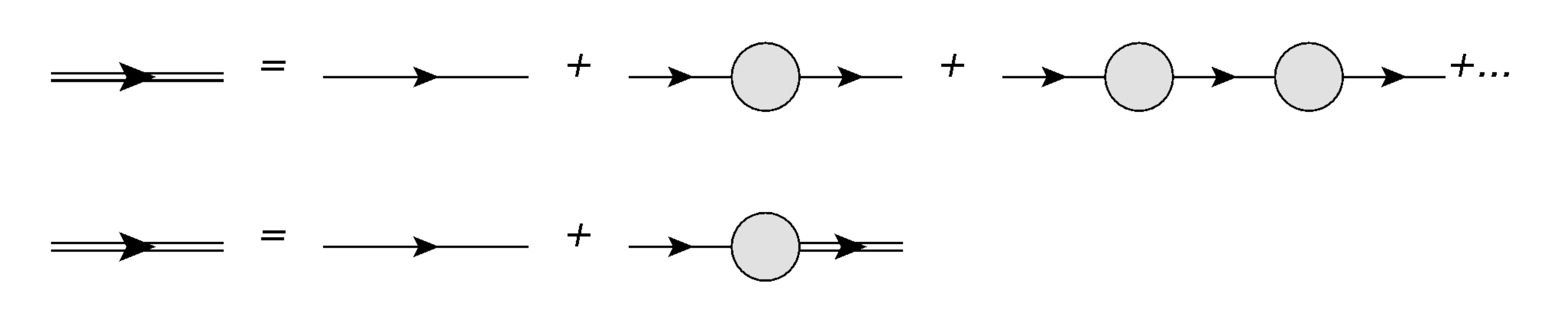}
  \caption{The diagrammatic representation of the Dyson's series. The figure shows that the exact fermion propagator can be obtained as the sum of the 
  radiative corrections. The double line represents the full fermion propagator, the grey blob is the self-energy which 
 incorporates the radiative corrections. It can be shown that this sum is a geometric series and has the algebraic form of 
  \eqref{dsser}.}
  \label{fig:DSer}
\end{figure}

Now that we have the formula for the self-energy, we need another equation which expresses the fermion propagator 
as a function of the self-energy. For this purpose Dyson's series can be used \cite{Pes}. It can be shown that 
summing up all the radiative corrections to the free propagator will lead us to a sum of a geometric series (see Fig.~
\ref{fig:DSer}). This results in an algebraic form:
\begin{equation}\label{dsser}
  {\cal G}^{-1}(p) = Z(p-m) -\Sigma(p),
\end{equation}
where we introduced $Z$ as the wave function renormalization factor coming from the vertex part. At this point we are 
not done yet, since in order to compute anything from \eqref{DSeq} and \eqref{dsser} we would need an 
approximation of the vertex function $\Gamma^{\mu}$ to close the system of equations. However, in the case of the 
BN model we can relate the vertex correction to the fermion propagator in a unique way. This is the key point of the 
computation presented in \cite{JakoMati1}, and it can be achieved using the Ward-Takahashi identities. This exact 
equation is coming from the current conservation analogously to the QED  case \cite{Pes,itzy}:
\begin{equation}\label{ward}
  k_\mu \Gamma^\mu(k;p-k,p) ={\cal G}^{-1}(p) - {\cal G}^{-1}(p-k).
\end{equation}
In this model, however, the vertex function is proportional to
$u^\mu$. In principle, the Lorentz-index in this model can come from
$u^\mu$ or from any of the momenta. But, since the fermion propagator
depends on the four-momentum in the form $u_\mu p^\mu$, the fermion-photon
vertex does not depend on the momentum components which are orthogonal to
$u^\mu$. Therefore the Lorentz-index which comes from $k^\mu$ in fact
comes from the longitudinal part of $k^\mu$, i.e. proportional to
$u^\mu$. So we can write $\Gamma^\mu(k;p-k,p) = u^\mu
\Gamma(k;p-k,p)$.
Using equations \eqref{DSeq}, \eqref{dsser} and \eqref{ward} we can compute the exact fermion propagator. The detailed calculation can 
be found in \cite{JakoMati1}, where the Feynman gauge is used and the UV renormalization is performed systematically. This all boils 
down to the to the following differential equation:
\begin{equation}
  (u p-m){\cal G}' + {\cal G} = -\frac{\alpha}{\pi} {\cal G}, 
\end{equation}
and its solution is:
\beq\label{fullsol}
 {\cal G}(p) \propto (u p-m)^{-1-\frac{\alpha}{\pi}}.
\eeq

This is indeed the exact solution for the fermion propagator in the BN model \cite{BN}. Comparing it to the free 
propagator, which has the form of $1/(up-m)$ the dressed solution has a power-law correction. This correction is due 
to the fully quantised soft radiation field, which shows in this respect that the electron cannot be thought as an 
independent object, but it is rather quasiparticle surrounded by the soft photon cloud. In the following we are going to give the solution for 
the finite temperature case. 

\subsection{The Bloch-Nordsieck model at $T>0$}

In this section we will show the solution for the BN model at finite temperatures. There are a few studies on the finite temperature BN 
model with different approximations \cite{IancuBlaizot1,IancuBlaizot2,Weldon:1991eg,Weldon:2003wp,Fried:2008tb}, however, the only 
analytically closed formula for the spectral function of the fermion is derived in \cite{JakoMati2}; we will present the results from the latter.
We already discussed that the $u^{\mu}$ parameter
as the fixed four-velocity of the fermion implies that the Bloch-Nordsieck model
describes the regime where the soft photon fields do not have energy
even for changing the velocity of the fermion (no fermion
recoil). This leads to the interpretation that the fermion is a hard
probe of the soft photon fields, and as such it is not part of the
thermal medium \cite{IancuBlaizot1, IancuBlaizot2}.\\ 
We are interested in the finite temperature fermion propagator. To
determine it, we use the real time or Keldysh formalism (for details, see
\cite{LeBellac}). Here the time variable runs over a contour
containing forward and backward running sections ($C_1$ and
$C_2$). The propagators are subject to boundary conditions which can
be expressed as the KMS (Kubo-Martin-Schwinger) relations \cite{LeBellac}. The physical time can be expressed 
through the contour time $t={\cal
  T}(\tau)$. This makes possible to work with fields living on a
definite branch of the contour, $\Psi_a(t,\x) = \Psi(\tau_a,\x)$ where
${\cal T}(\tau_a)=t$, and $\tau_a\in C_a$ for $a=1,2$; and similarly
for the gauge fields. The propagators are matrices in this notation, in particular the fermion and the photon propagator, 
respectively, reads as:
\bea
  i{\cal G}_{ab}(x)&=&\exv{T_{\cal C} \Psi_a(x) \Psi_b^\dagger (0)},\nn
  iG_{\mu\nu,ab}(x)&=&\exv{T_{\cal C} A_{\mu a}(x) A_{\nu b} (0)},
\eea
where $T_{\cal C}$ denotes ordering with respect to the contour variable
(contour time ordering). For a generic propagator $G_{11}$ corresponds to the Feynman
propagator, and, since the $C_2$ contour times are always larger than
the $C_1$ contour times, $G_{21}=G^>$ and $G_{12}=G^<$ are the
Wightman functions. The KMS relation for a bosonic/fermionic
propagator reads $G_{12}(t,\x)= \pm G_{21}(t-i\beta,\x)$ which has the
following solution in Fourier space (with $k=(k^0,{\bf k})$)
\bea
  \label{eq:id1}
  iG_{12}(k) &= &\pm n_\pm(k_0) \rh(k),\nn
  iG_{21}(k) &=& (1\pm n_\pm)(k_0) \rh(k),
\eea
where 
\begin{equation}
  n_\pm(k_0) = \frac1{e^{\beta k_0}\mp 1}\quad \mathrm{and}\quad \rh(k)
  = iG_{21}(k)-iG_{12}(k)
\end{equation}
are the distribution functions (Bose-Einstein (+) and Fermi-Dirac (-)
statistics), and the spectral function, respectively. It is sometimes
advantageous to change to the R/A formalism with field assignments
$\Psi_{1,2} = \Psi_r\pm\Psi_a/2$. Then, one has $G_{aa}=0$ for both the
fermion and the photon propagators. The relation between the propagators
in the Keldysh formalism and the R/A propagators reads
\begin{equation}
  \label{eq:id2}
  G_{rr}=\frac{G_{21}+G_{12}}2,\quad G_{11} = G_{ra} + G_{12},\quad
  \rh=i G_{ra}-iG_{ar}.
\end{equation}
The $G_{ra}$ propagator is the retarded, the $G_{ar}$ is the advanced
propagator, $G_{rr}$ is usually called the Keldysh propagator in the framework of 
the R/A formalism (not to confuse with the propagators {\emph  {in}} the Keldysh formalism).\par
At zero temperature, as we could see in the previous section, 
the fermionic free Feynman propagator in the BN model has a 
single pole which means that there are no antiparticles in the model and it coincides with the retarded propagator of the theory.
Consequently, all closed fermion loops are zero, thus there is
no self-energy correction to the photon propagator at zero
temperature, as we already discussed it.
So, we will set ${\cal G}_{12}=0$, therefore the closed fermion loops as well as the photon self-energy
remain zero even at finite temperature. Another, mathematical reason,
why we must not consider dynamical fermions -- which could show up in
fermion loops -- is that the spin-statistics theorem
\cite{Pes} forbids a one-component dynamical fermion
field.\\
This means that now the \emph{exact} photon propagator reads in
Feynman gauge at finite temperature
\beq
  G_{ab,\mu\nu}(k) = -g_{\mu\nu} G_{ab}(k),\nn
  G_{ra} = \frac1{k^2}\biggr|_{k_0\to k_0+i\ep}.
\eeq
And the \emph{exact} photon spectral function is
\beq
\rh(k)= 2\pi\sgn(k_0) \delta(k^2).
\eeq
All other propagators can be expressed using the identities in \eqref{eq:id1}
and in \eqref{eq:id2}.
\par
\begin{figure}
\begin{subfigure}{.5\textwidth}
\centering
  \includegraphics[width=.65\linewidth]{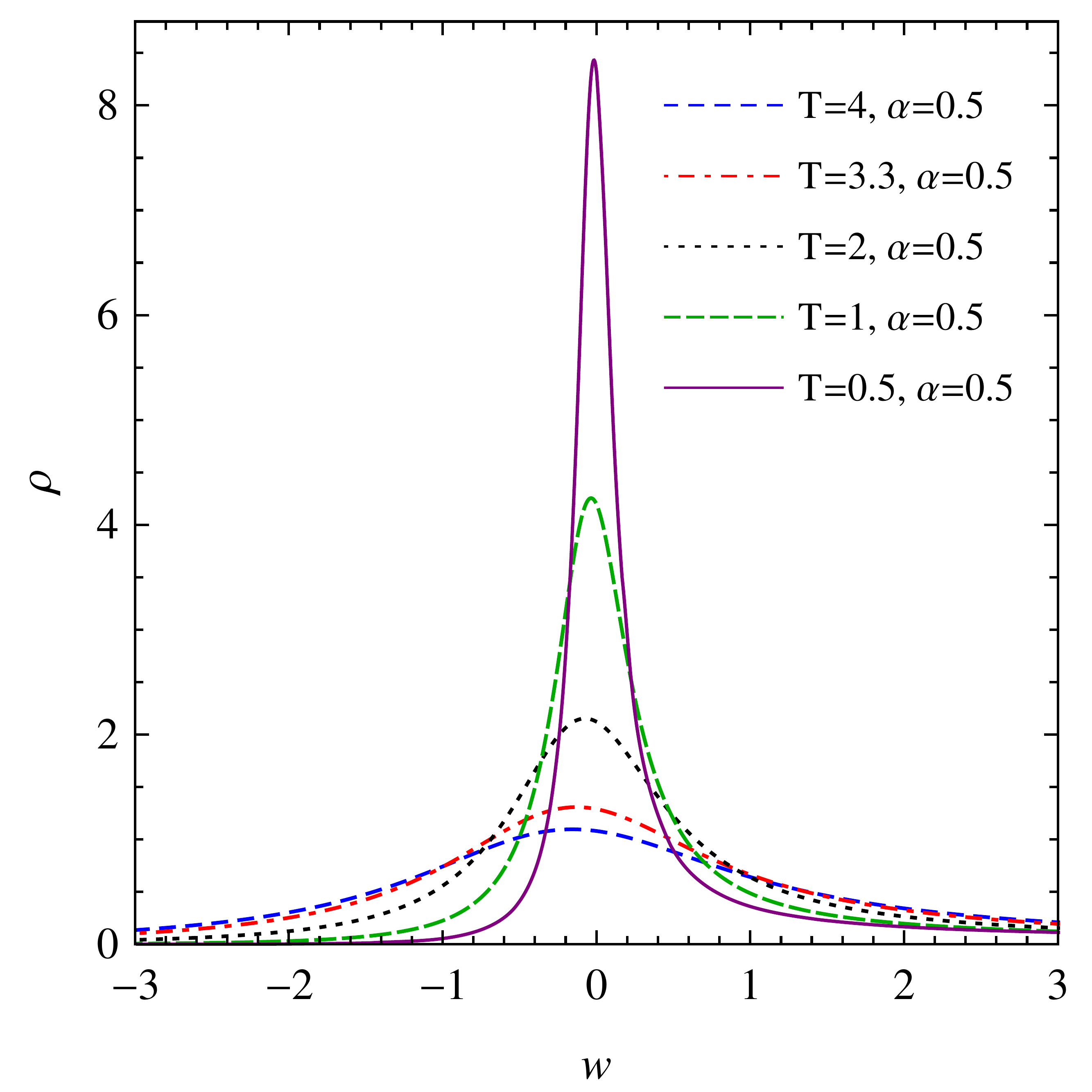}
  \caption{ }
  \label{fig:sfig1}
\end{subfigure}%
\begin{subfigure}{.5\textwidth}
 \centering
  \includegraphics[width=.65\linewidth]{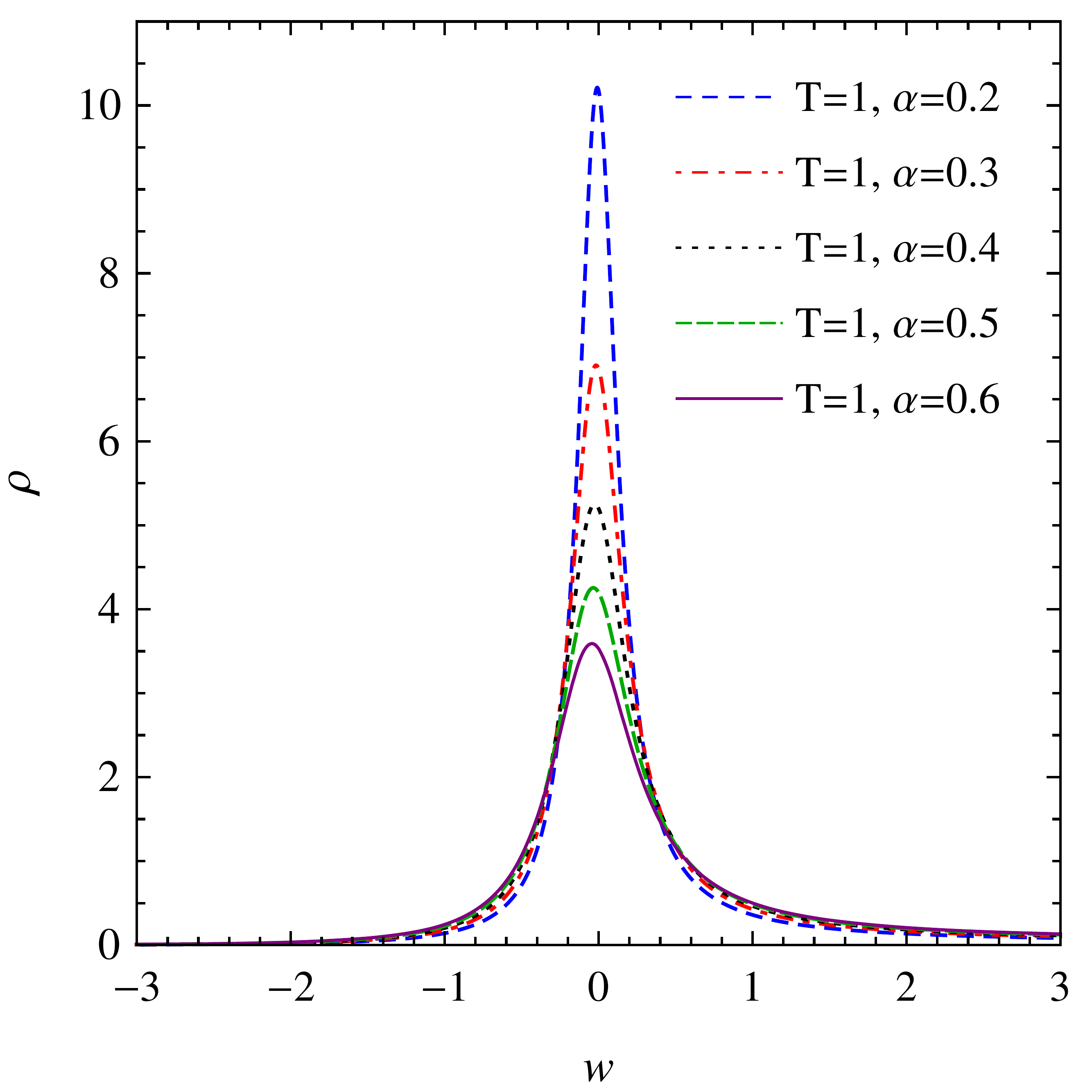}
  \caption{ }
  \label{fig:sfig2}
\end{subfigure}\vspace{.2cm}
\centering
\begin{subfigure}{.51\textwidth}
\centering
\includegraphics[width=.65\linewidth]{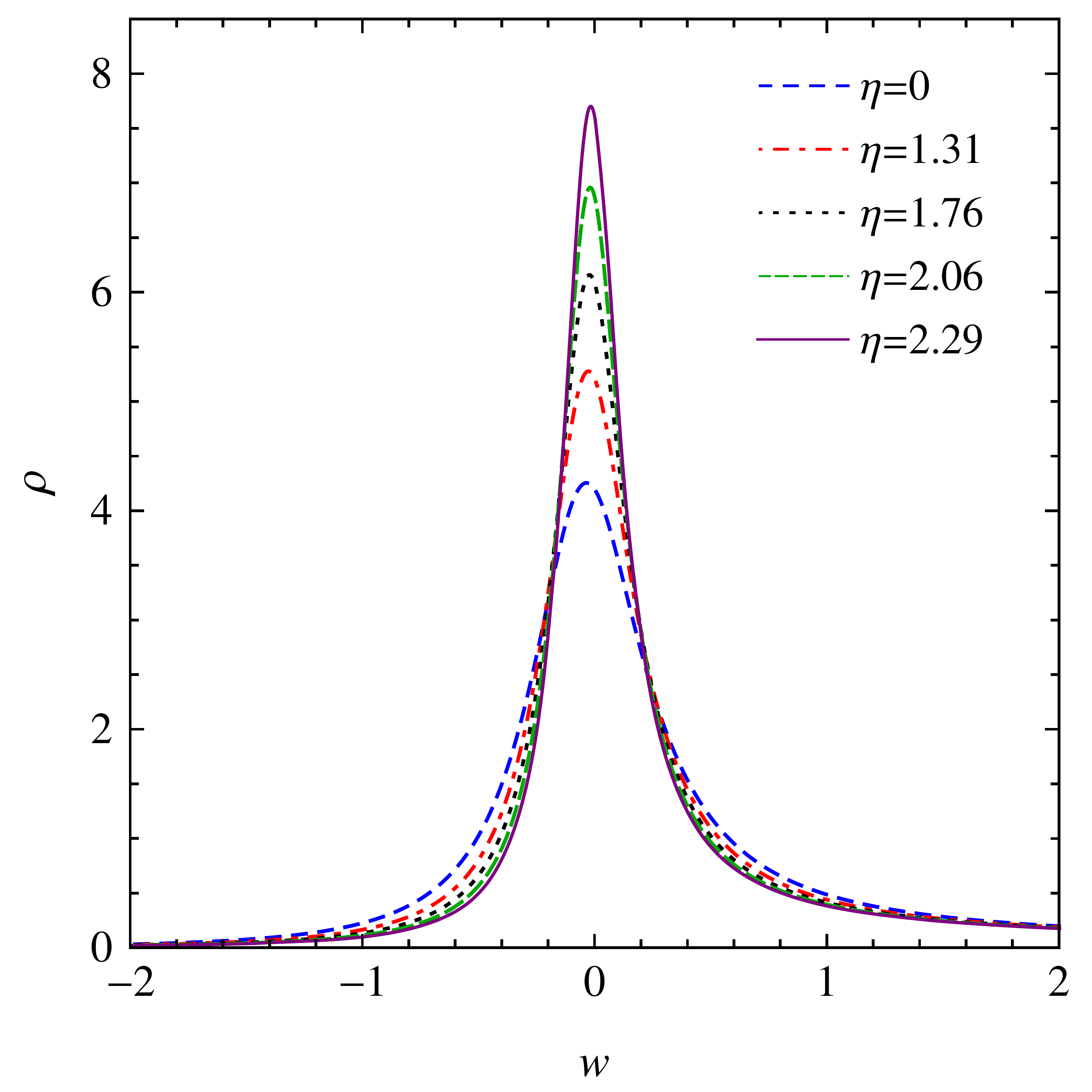} 
  \caption{ }
  \label{fig:sfig3}
\end{subfigure}
\caption{The fermionic spectral functions in the BN model at finite temperature. In (a) and (b) we see the broadening 
of the quasiparticle peak as the temperature and the coupling constant increased at fixed $\alpha=0.5$ and 
temperature $\beta=1/T=1$, respectively. In (c) the velocity dependence of the spectral function is shown for $
\alpha=0.5$. The $\eta$ values are rapidities, $|{\bf{v}}|=\tanh\eta$. The peak region becomes sharper with increasing 
$\eta$ ($|{\bf{v}}|$).}
\label{fig:spectal}
\end{figure}

 Now we can consider the system of self-consistent equations at finite temperatures. The derivation of the DS 
equations at non-zero temperature can be found in \cite{JakoMati2} where the CTP (Closed Time 
Path) formalism was applied (see \cite{LeBellac}). 
Thus, one can express the equation for the fermion self-energy with the two-component notation as
it can be seen in Fig.~\ref{fig:DS}.
In terms of analytic formulas it reads:
\begin{equation}
  \Sigma_{ab}(x,y) = i\alpha_a e^2u_\mu \sum_{c,d=1}^2 \int\! d^4w
  d^4z\, {\cal G}_{ac}(x,w) G^{\mu\nu}_{ad}(x,z) \Gamma_{\nu;dcb}(z;w,y),
\end{equation}
where $\alpha_a=(-1)^{a+1}$. In Fourier space it is:
\begin{equation}\label{fintdyson}
  \Sigma_{ab}(p) = i\alpha_a e^2u_\mu \sum_{c,d=1}^2 \pint 4k {\cal
    G}_{ac}(p-k) G^{\mu\nu}_{ad}(k) \Gamma_{\nu;dcb}(k;p-k,p).
\end{equation}
This equation is the non-zero temperature equivalent of \eqref{DSeq}, the only difference here is that,
since we evaluate each operator on a given time contour, we need to indicate them, thus we use the lower
indices for this purpose.\\
The vertex function in this case can be derived in a similar manner, too. The fact that the 
vertex function is proportional to $u^{\mu}$ is crucial again, in order to be able to apply the 
Ward-identities in a way we did already in the zero temperature computation. The derivation of Ward-identities at finite
temperature is a straightforward generalisation of the formula that we have at zero temperature.
It is easy to rewrite it in the two-component formalism, taking
into account that to satisfy the delta functions requirement the time arguments
must be on the same contour. One finds in Fourier space
\begin{equation}
  k_\mu \Gamma^\mu_{abc}(k;p,q) = \left[ \delta_{ab} {\cal
      G}_{bc}^{-1}(q) - \delta_{ac} {\cal G}_{bc}^{-1}(p) \right] \,
  (2\pi)^4 \delta(k+p-q).
\end{equation}
In the BN model, because of the special property of the
vertex function, it is \emph{completely} determined by the fermion propagator in the form:
\begin{equation}
  \label{eq:vertex}
  \Gamma_{abc}(k;p,q) = \frac1{uk}\left[  \delta_{ab} {\cal
      G}_{bc}^{-1}(q) - \delta_{ac} {\cal G}_{bc}^{-1}(p)\right]
  \biggr|_{p=q-k},
\end{equation}
therefore the DS equations for the fermion propagator become closed. As the third component of the 
system of equations we will use \eqref{dsser} again. From \eqref{dsser}, \eqref{fintdyson} and \eqref{eq:vertex} the 
detailed calculation of the full solution can be found in \cite{JakoMati1}. At the end of the procedure we will get the 
following expression for the spectral function (defined as the discontinuity of the retarded propagator) in the special 
reference frame where $u=(1,0,0,0)$:
\begin{equation}
  \label{eq:exactsol}
  \rho(w) = \frac{\displaystyle N_\alpha \beta \sin\alpha\,e^{\beta w/2}}
  {\displaystyle\cosh(\beta w)-\cos\alpha}\, \frac 1{\displaystyle
    \left|\Gamma\left(1+\frac{\alpha}{2\pi} + i\frac {\beta
          w}{2\pi}\right)\right|^2},
\end{equation}
where $N_\alpha$ is an $\alpha$ dependent normalization factor, $\beta$ is the inverse temperature and $w=p_0-m$ 
is the variable (the energy shifted by the mass parameter). It can be shown that  this result is completely consistent 
with the zero temperature one when one takes the $T\to0$ limit \cite{JakoMati2}. In Fig.~\ref{fig:spectal} we show the spectral 
function for different parameters of the temperature and the coupling constant $\alpha$. By increasing the temperature the 
broadening of the quasiparticle peak can be observed and the same effect can be achieved by increasing the coupling of the 
theory. The halfwidth of the curve can be related to the inverse of the lifetime of the quasiparticle; hence increasing the 
temperature lowers the lifetime of the quasiparticle which is physically sensible. The same can be said about varying the 
coupling constant. Since the Lorentz invariance does not hold at finite temperature we need to calculate the spectral function 
numerically when one goes from the rest frame of the fermion to a different one, for details see \cite{JakoMati2}. By increasing 
the four-velocity of the fermion the halfwidth shrinks, indicating more stable quasiparticles.
The finite temperature fermion propagator can be retained using the spectral function in \eqref{eq:exactsol}:
\beq
{\mathcal G}(p)=\int \limits_{0}^{\infty} \frac{d\omega}{2\pi} \frac{\rho(\omega)}{p-\omega+i\epsilon}.
\eeq
Since the $T\to 0$ limit of \eqref{eq:exactsol} gives back the spectral function that can be obtained for zero temperature 
(c.f. \cite{JakoMati1}), this limit of the finite temperature propagator also gives back the $T=0$ formula \eqref{fullsol} for this limit.

\section{Applications}
\begin{figure}[htbp]
  \centering
  \includegraphics[width=0.7\textwidth]{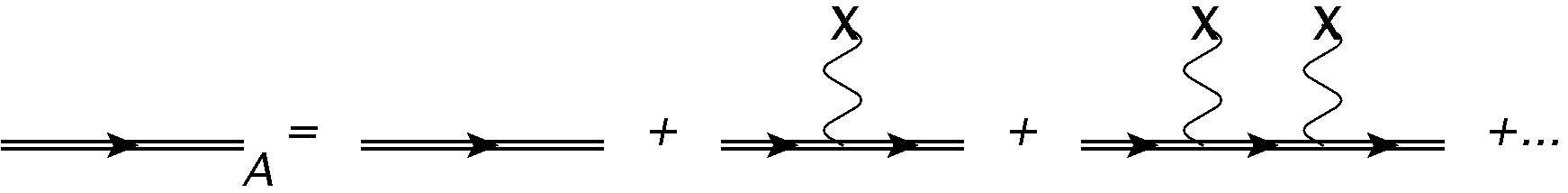}
  \caption{The BN propagator in the presence of an external field. The double line denotes the BN propagator, the letter "X" is 
  for the classical source producing the external field which are the wavy lines. The subscript $A$ on the left hand side denotes that the 
  propagator is in the external field.}
  \label{fig:ext}
\end{figure}

In this section we will present some of the ideas of the possible applications. As we have demonstrated that the soft photon 
contribution to the propagator of the electron is important both for $T=0$ and $T>0$ [\eqref{fullsol} and 
\eqref{eq:exactsol}] we can use these expressions when we calculate processes in external fields. In such cases we 
can use the usual formula for the propagation in external field (see Fig.~\ref{fig:ext}), except we use the propagator 
obtained from the BN model instead of the free one (both for $T=0$ and $T>0$):
\bea
{\mathcal G}_A(x_f,x_i)&=&{\mathcal G}(x_f,x_i)+\int d^4 x_1 {\mathcal G}(x_f,x_1) e \slashed A(x_1) {\mathcal 
G}(x_1,x_i)\nn 
&&+ \int \int d^4 x_1 d^4 x_2 {\mathcal G}(x_f,x_1)e \slashed A(x_1){\mathcal G}(x_2,x_1)e \slashed A(x_2)
{\mathcal G}(x_2,x_i)+...
\eea
Here we used ${\mathcal G}_A$ for the full propagator in the external field $A^\mu$ and ${\mathcal{G}}$ for the BN 
fermion propagator.
Since we have a simpler form of the BN propagator in momentum space we should use for our calculations:
\bea
{\mathcal G}_A(p_f,p_i)&=&{\mathcal G}(p_f)(2\pi)^4\delta^4(p_f-p_i)+\int \, \frac{d^4 p_1}{(2\pi)^4} {\mathcal G}(p_f) e \slashed A(p_1) {\mathcal G}(p_i)\nn 
&&+ \int \int \frac{d^4 p_1}{(2\pi)^4} \, \frac{d^4 p_2}{(2\pi)^4} \, {\mathcal G}(p_f)e \slashed A(p_1){\mathcal G}(p_i
+p_2)e \slashed A(p_2){\mathcal G}(p_i)(2\pi)^4\delta^4(p_f-p_1-p_2-p_i)+...
\eea
In this expression $A^\mu$ can be considered as an arbitrary external field. For example we can use the Coulomb potential generated by 
a static point charge $-Z e$:
\beq
A_0(x)=-\frac{Z e}{|{\bf{x}}|}, \qquad {\bf{A}}(x)={\bf{0}}.
\eeq
Or in momentum space $A_0=-Z e 4\pi/|{\bf q}|$. Alternatively, we can also use the screened Coulomb field: $A^{\mu}
=-Ze\delta^{\mu \,0}/|{\bf{x}}| \exp(-|{\bf{x}}|/l)$, where $l$ is the screening length.

The interaction of the dressed electron with external field could have an application in the computation of processes 
like the multiphoton Compton scattering and the second order processes like the Bremsstrahlung, which are getting 
more important as the laser intensities are increasing. The main idea is similar to those of used in \cite{jent,pia} 
but instead of the Volkov propagator of the fermion the Bloch-Nordsieck propagator could be used.  

\section{Summary}
In this paper we discussed the infrared catastrophe in the framework of QED and its resolution using the Bloch-
Nordsieck summation of radiative corrections. We sketched an alternative derivation of 
the Bloch-Nordsieck propagator with the aid of the Ward-identities; the full derivation is presented in \cite{JakoMati1} and for 
finite temperatures in \cite{JakoMati2}. The Bloch-Nordsieck propagator describes the propagation of the dressed electron in the deep 
infrared regime of the QED, where the soft photon contributions can be completely summed up. Speculation of the 
possible applications of the results are given in the last section. The detailed calculations of these processes are topic of current research projects.


\section{Acknowledgement}
The ELI-ALPS project (GOP-1.1.1-12/B-2012-0001) is supported by the European Union and co-financed by the European Regional 
Development Fund.

\end{document}